\documentclass[%
 aip,
 amsmath,amssymb,
reprint,%
]{revtex4-1}

\usepackage{graphicx}
\usepackage{dcolumn}
\usepackage{bm}

\usepackage[utf8]{inputenc}
\usepackage[T1]{fontenc}
\usepackage{mathptmx}

\usepackage[dvipsnames]{xcolor}
\usepackage{graphicx}
\usepackage{dcolumn}

\usepackage{calrsfs}
\DeclareMathAlphabet{\mathcal}{OMS}{zplm}{m}{n}
\usepackage{placeins} 
\usepackage{soul}
\usepackage{enumerate} 

\newcommand{\dey}{\partial_{y} }

\definecolor{amber}{rgb}{1.0, 0.75, 0.0}
\newcommand{\ud}{\mathop{}\!\mathrm{d}}  

\begin{document}

\preprint{AIP/123-QED}

\title[]{Wave focusing and related multiple dispersion transitions in plane Poiseuille flows}

\author{F. Fraternale}
\altaffiliation[Author's affiliation at the time of the present analyses: ]{Dipartimento di Scienza Applicata e Tecnologia, Politecnico di Torino, Torino, Italy 10129.}
\affiliation{ 
Center for Space Plasma and Aeronomic Research, The University of Alabama in Huntsville, Huntsville AL 35805, USA
}%
\author{G. Nastro}%
\altaffiliation[Author's affiliation at the time of the present analyses: ]{Dipartimento di Scienza Applicata e Tecnologia, Politecnico di Torino, Torino, Italy 10129.}
\affiliation{ 
ISAE-SUPAERO, Universit\'e de Toulouse, Toulouse, France 31055
}%

\author{D. Tordella}
\email[Author to whom correspondence should be addressed: ]{daniela.tordella@polito.it}
\affiliation{%
Dipartimento di Scienza Applicata e Tecnologia, Politecnico di Torino, Torino, Italy 10129
}%


\begin{abstract}
Motivated by the recent discovery of a dispersive-to-nondispersive transition for linear waves in shear flows, we accurately explored the wavenumber-Reynolds number parameter map of the plane Poiseuille flow, in the limit of least-damped waves. We have discovered the existence of regions of the map where the dispersion and propagation features vary significantly from their surroundings. These regions are nested in the dispersive, low-wavenumber part of the map. This complex dispersion scenario demonstrates the existence of linear dispersive focusing in wave envelopes evolving out of an initial, spatially localized, three-dimensional perturbation.  An asymptotic wave packet's representation, based on the saddle-point method, allows to enlighten the nature of the packet's morphology, in particular the arrow-shaped structure and spatial spreading rates. A correlation is also highlighted between the regions of largest dispersive focusing and the regions which are most subject to strong nonlinear coupling in observations.\par\vskip10pt

This article may be downloaded for personal use only. Any other use requires prior permission of the author and AIP Publishing. This article appeared in Physics of Fluids 33, 034101 (2021) \url{https://doi.org/10.1063/5.0037825} and may be found at \url{https://aip.scitation.org/doi/full/10.1063/5.0037825}
\end{abstract}

\maketitle


\section{\label{sec1:intro}Introduction}

Dispersion is a fundamental property of traveling waves, and its terminology stems from types of solutions rather than types of governing equations, as well explained in the 1974 monography by \citet{whitham1974} dedicated both to linear and nonlinear waves. It is commonplace to talk about \textit{dispersive equations}: well-known examples of linear and nonlinear dispersive partial differential equations are the \textit{Airy}, \textit{Euler--Bernoulli beam}, \textit{Klein--Gordon},  \textit{Schr{\"o}dinger}, \textit{Korteweg--de Vries} and  \textit{Boussinesq} equations. The reader can find general information in Refs. \onlinecite{sulem1991book,craik1985, karpman2016}. 

Dispersive wave focusing, i.e. the time-space localization of wave-train energy, is frequently encountered in physical sciences in very diversified areas. This mechanism relies on the phase modulation of perturbation wave-trains and produces regions in which disturbances can ``focus'' and reach finite amplitudes. Mention can be made of two examples: surface waves on water of finite depth, where dispersive focusing was suggested as a giant wave generation mechanism \cite{ablowitz1979,dysthe1999,henderson1999,onorato2001,slunyaev2002}, and wave-guides in integrated optical circuits \cite{smit1988}.  In particular, in the field of water surface waves, the new knowledge of propagation pairing of long waves with short waves, and the interplay of their angle of inclination, introduces a new interpretation tool into the analysis of both linear and nonlinear wave interactions. This dynamical aspect has not been considered in great detail in the study of turbulence transition. To date, dispersive focusing has not yet been reported in the field of perturbation waves traveling in non-stratified bounded flows within the framework of the Navier--Stokes equations.

In these flows, propagating waves are at the root of fluid flow instability and transition to turbulence. In particular, the growth of wave packets or localized spots is of great interest in the subcritical route to turbulence, also known as bypass transition \cite{morkovin1969}. In past literature, the propagation and dispersion features of such internal waves has attracted less attention than the transient mechanisms responsible for their potential amplification, a process which was hitherto considered as the cause of transition to turbulence and, in the last few decades, has been framed within theories based on non-normal growth [see, e.g., Refs. \onlinecite{ttrd1993,schmid2007}].

In the past, many authors have devoted attention to such perturbation waves in shear flows, whilst no accurate characterization of their dispersion properties has been conducted so far. Linear and nonlinear wave dispersive focusing may be a potential agent of the catastrophic transition to turbulence observed for larger Reynolds numbers than the transitional thresholds beyond which uniform turbulence is observed, instead of the coexistence of laminar and turbulent patches. In fact, the morphology of wave packets in shear flows has been mainly described through information concerning the global structure, deduced either from laboratory experiments \cite{emmons1951,carlson1982,alavyoon1986,klingmann1992,lemoult2013,lemoult2014, klotz2017a,klotz2017b} or from Navier-Stokes direct numerical simulations (DNSs) \cite{lundbladh1991,henningson1991,duguet2012}, where it is not easy to pick out the dynamics of the individual wave component. 

The present study deals with a recently discovered instance of the complexity of dispersion properties for linear waves traveling within viscous, incompressible fluid flows governed by the Navier--Stokes equations under linearized dynamics. We intend to show that this uneven scenario in wave dispersion features allows to gather significant information about the morphology and propagation of localized, three-dimensional (3-D) disturbances in bounded flows. In particular, the system we consider is the planar Poiseuille flow (PPF) between two infinitely long, parallel plates at a fixed distance 2$h$ apart, as sketched in Fig. \ref{fig:flowscheme}. This flow is
driven by a pressure gradient in the flow direction, and is retarded by viscous drag along both plates, so that these forces are balanced.

Previously, we demonstrated the existence of a sharp dispersive-to-non-dispersive transition in PPF and wake flows for (normalized) wave number values near unity \cite{desanti2016,fraternale2018,fraternale_phdthesis}. This was done by numerically computing the long-term dispersion relation of the Orr-Sommerfeld (OS) and Squire eigenvalue problem \cite{o1907a,squire1933} within a four-decade range of the Reynolds number ($\mathcal{R}$) and a three-decade range of wavenumbers ($k$). Here, a detailed computation has been carried out for the phase velocity and the local group velocity of least-damped OS modes, extending the previous analyses in the limit of small wavenumbers. We adopt the generalization of the concept of the group velocity given by \citet{whitham1974} for wave packets in purely dispersive homogeneous media to the case of dispersive and dissipative media \cite{muschietti1993, sonnenschein1998, gerasik2010}.

We will first present newly discovered properties of wave dispersion in PPF. These features appear as transitions of the dispersion nature of the least-damped OS mode in the $\mathcal{R}-k$ map (hereinafter also referred to as \emph{dispersion map}). In particular, we show the existence of several regions in the small-wavenumber portion of this map where wave dispersion characteristics change significantly from the surroundings, and the nature of the least-damped OS mode changes as well.

Then, we show that this picture produces linear dispersive focusing. In fact, when a wave packet is assembled via superposition of monochromatic waves ranging from the smallest to the largest wavenumber considered here, it occurs both that components having similar wavenumbers can propagate with different speeds, which yields dispersion, and, on the other hand, waves with distinct wavelength can show very similar propagation features. 
Via a simple propagation scheme \cite{gaster1965}, we highlight the existence of multiple loci for dispersive focusing in the physical space where the packet propagates.  Since dispersive effects are known to play a leading role in pattern formation and wave dynamics [see, e.g., Ref. \onlinecite{craik2005}], this new propagation scenario can help to understand the morphology of perturbation clusters and wave packets, at least in their early evolution before nonlinear effects occur, triggering secondary flow bifurcations that are not predicted by the linear approach. In fact, we show that the described focusing explains the major features of the morphology and propagation of localized, 3-D disturbances (or \emph{spots}) in channel flows, such as the arrow-shaped structure, the leading streaks, and the trailing waves at the spot's wingtips. The comparison of our results with laboratory experiments also suggests that wave focusing in the early linear phase may play  an important role in the onset of nonlinear coupling and consequent transition to turbulence in shear flows. This topic will need to be further investigated in a future study.

The paper is organised as follows. In Sec. \ref{sec2:math}, we recall the physical problem and the mathematical model. The results concerning the dispersion relation of PPF are discussed in Sec. \ref{sec3:disp_maps}. The unsteady evolution of localized wave packets and their asymptotic representation are presented in Sec. \ref{sec4:wp_asymp_repres}, and conclusions are drawn in Sec. \ref{sec5:conclusions}. Appendix \ref{app:A} is devoted to the technical details about our numerical simulations. 

\section{\label{sec2:math}Physical problem and mathematical framework}

\begin{figure}[ht!]
    \includegraphics[width=\columnwidth]{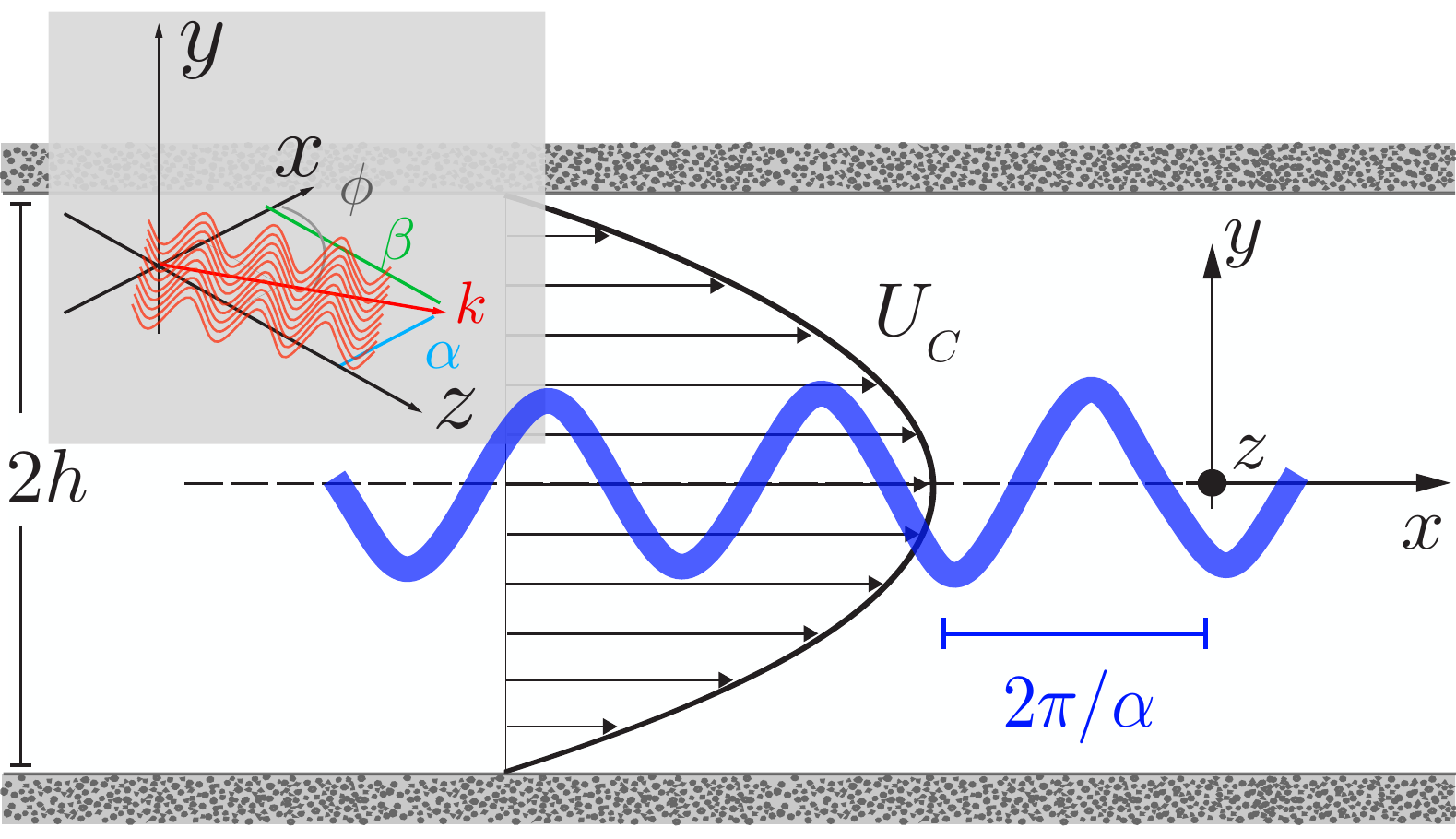}\vskip-5pt
	\caption{Sketch of the plane Poiseuille flow, coordinate system and 3-D wave perturbation. The reference velocity is the centerline velocity $U_{C}$, and the reference length is the half thickness of the channel,  $h$. The Reynolds number is then $\mathcal{R} = h U_{C} / \nu$, where $\nu$ is the kinematic viscosity of the fluid. The planar wavenumber vector and the wave angle of the perturbation are respectively $\bm{k}=\alpha \bm{e_x}+\beta \bm{e}_z$ and $\phi=\tan^{-1}(\beta/\alpha)$.}
	\label{fig:flowscheme}	
\end{figure}

We consider the plane Poiseuille flow (PPF), where inertia and molecular diffusion are the two only players. Figure \ref{fig:flowscheme} presents a longitudinal cut of the channel, and reports the coordinate system and the reference quantities used for normalization. 
We consider as characteristic length and velocity scales the half height of the channel $h$ and the centerline velocity $U_{C}$, respectively. As we address incompressible buoyancy-free flows, the governing equations are the Navier-Stokes equations which read, in dimensionless form:
\begin{gather}
    \boldsymbol{\nabla} \boldsymbol{\cdot} \bm{u} = 0
    \label{eq:continuity}\\
    D_t \bm{u} = - \boldsymbol{\nabla} p + \frac{1}{\mathcal{R}} \boldsymbol{\Delta} \bm{u} \,,
    \label{eq:momentum}
\end{gather}
where $D_t=\partial_t+(\bm{u} \boldsymbol{\cdot} \boldsymbol{\nabla})$ denotes the material derivative and $ \mathcal{R}  = h U_{C} / \nu$ is the Reynolds number with $\nu$ the fluid kinematic viscosity. No-slip boundary conditions are imposed at the walls ($\bm{u}=\bm{0}$ at $y=\pm 1$).
The parallel basic flow $\bm{U}= U(y) \bm{e_x}$ consists of the well known parabolic profile of Poiseuille, stationary solution of Eqs. (\ref{eq:continuity}-\ref{eq:momentum}):
\begin{equation}
   U(y) = 1 - y^2 \,.
    \label{eq:Poiseuille_profile}
\end{equation}
The governing equations for a small perturbation $\widetilde{\bm{u}}=\bm{u}-\bm{U}$, $\widetilde{p}=p-P$,  are obtained by linearizing Eqs. (\ref{eq:continuity}-\ref{eq:momentum})  about the basic flow (\ref{eq:Poiseuille_profile}): 
\begin{gather}
    \boldsymbol{\nabla} \boldsymbol{\cdot} \widetilde{\bm{u}} = 0
    \label{eq:continuity_p}\\
    \partial_t \widetilde{\bm{u}} + \bm{U} \boldsymbol{\cdot} \boldsymbol{\nabla} \widetilde{\bm{u}} + \widetilde{\bm{u}} \boldsymbol{\cdot} \boldsymbol{\nabla} \bm{U} = - \boldsymbol{\nabla} \widetilde{p} + \frac{1}{\mathcal{R}} \boldsymbol{\Delta} \widetilde{\bm{u}} \,.
    \label{eq:momentum_p}
\end{gather}
We resort to both a modal and non-modal approach in order to determine linear wave disturbances that are likely to grow over the Poiseuille's profile. Taking into account the streamwise and spanwise homogeneity of the base flow (\ref{eq:Poiseuille_profile}), a general perturbation $\widetilde{\bm{q}}=[\widetilde{\bm{u}},\widetilde{p}]=[\widetilde{u},\widetilde{v},\widetilde{w},\widetilde{p}]$ can be expressed as a Fourier integral,
\begin{equation}
\widetilde{\bm{q}}(\bm{x},t)=\int_{-\infty}^{+\infty} \hat{ \bm{q}}(\bm{k};y,t) e^{i\bm{k}\cdot\bm{x}}\ud\bm{k}\,,
\label{eq:fourierintegral}
\end{equation}
where $\bm{k}=\alpha \bm{e_x}+\beta \bm{e}_z$ is the planar wavenumber vector of streamwise and spanwise components, $\alpha$ and $\beta$, respectively. The wave angle is defined as $\phi=\tan^{-1}(\beta/\alpha)$ [see Fig. \ref{fig:flowscheme}].

We denote $k$ the magnitude of the planar wavenumber vector and this convention will be used throughout the paper. By using the wall-normal velocity-vorticity formulation \cite{criminale_book}, the linearized non-modal equations  in the wavenumber space read:
\begin{gather}
    \left[ ( \partial_t + i \alpha U) (\partial_y ^2 - k^2) - i \alpha U'' - \frac{1}{\mathcal{R}}(\partial_y ^2 - k^2)^2\right] \hat{v} = 0
    \label{eq:wall_normal_vel}\\
    \left[ ( \partial_t + i \alpha U) - \frac{1}{\mathcal{R}}(\partial_y ^2 - k^2)\right] \hat{\omega}_y = -i \beta U' \hat{v} \,,
    \label{eq:wall_normal_vort}
\end{gather}
where $\tilde{v}$ and $\tilde{\omega}_y$ are, respectively, the wall normal velocity and vorticity and the prime symbol stands for a derivation with respect to the wall-normal direction, $y$. The streamwise and spanwise velocity components can be easily recovered \emph{a posteriori} from $\hat{v}$ and $\hat{\omega}_y$. The following expressions are obtained from the continuity equation (\ref{eq:continuity_p}) and from the definition of the velocity curl:
\begin{gather}
    \hat{u} = \frac{i}{k^2} \left( \alpha \partial_y \hat{v} - \beta \hat{\omega}_y \right)
    \label{eq:streamwise_vel}\\
    \hat{w} = \frac{i}{k^2} \left( \beta \partial_y \hat{v} + \alpha \hat{\omega}_y \right) \,.
    \label{eq:spanwise_vel}
\end{gather}
The boundary conditions associated with the linearized system [Eqs. (\ref{eq:wall_normal_vel}-\ref{eq:wall_normal_vort})] are homogeneous and correspond to the \textit{no-slip} condition:
\begin{equation}
    \hat{v} (\pm 1) = \partial_y \hat{v} (\pm 1) = \hat{\omega}_y (\pm 1) = 0 \,.
    \label{eq:boundary_conditions }
\end{equation}
These boundary conditions are exactly satisfied by the Chandrasekhar-Reid functions used by our numerical method.  Note that different basic flow profiles, such as boundary layers, wakes, or jets, may accept different boundary conditions (e.g. the exponential decay can be imposed at $|y|\to\infty$). The Fourier representation of Eq. (\ref{eq:fourierintegral}) implies periodic boundary conditions at the $x$ and $z$ axes. In our numerical simulations described below, the domain is large enough to prevent boundary effects on the solution.

\section{\label{sec3:disp_maps}Long-term asymptotics of small-amplitude waves}

\subsection{\label{subsec3:modal_approach}Modal analysis approach}

The long-term asymptotic limit ($t\to\infty$) is obtained by imposing to Eqs. (\ref{eq:wall_normal_vel}-\ref{eq:wall_normal_vort}) the harmonic structure for the perturbations, i.e. $\partial_t \to -i \sigma$, where $\sigma(\bm{k}; \mathcal{R})=\sigma_r(\bm{k}; \mathcal{R})+i \sigma_i(\bm{k}; \mathcal{R})$ is a complex OS eigenvalue, defining the dispersion relation for the corresponding eigenmode. Let us recall that for bounded flows, the OS spectrum has infinitely many discrete eigenvalues. The dispersion relation $\sigma(\bm{k};\mathcal{R}) \equiv \omega(\bm{k}; \mathcal{R}) = \omega_r + i \omega_i$ is computed by solving numerically the eigenvalue problem associated with Eqs. (\ref{eq:wall_normal_vel}-\ref{eq:wall_normal_vort}). We adopt a fifth-order accurate Gal\"erkin method based on Chandrasekhar-Reid functions \cite{chandrasekhar61}, described in details in Refs. \onlinecite{desanti2016,fraternale_phdthesis}. The modal analysis approach yields the complex angular frequency whose real part $\omega_r$ allows to define the phase velocity
\begin{equation}
    \bm{c} \equiv \frac{\omega_r}{k} \, \bm{e}_k \,,
\end{equation}
where $\bm{e}_k = \cos\left(\phi\right) \bm{e}_x + \sin\left(\phi\right) \bm{e}_z $ stands for the direction cosine of planar wavenumber vector $\bm{k}$. The real group velocity is computed via numerical derivative from the kinematic definition
\begin{equation}
  \bm{v}_\mathrm{g}\equiv\bm{\nabla}_{k} \omega_r \,,
  \label{eq:complex_group_velocity}
\end{equation}
where $(\bm{\nabla}_{k})_i\equiv\partial / \partial k_i$ is the gradient in the wavenumber space. 
The corresponding imaginary part $\bm{\nabla}_{k} \omega_i$ was also computed and the discussion on its role will be resumed later in Sec. \ref{sec4:wp_asymp_repres}. A dispersion factor, $\bm{f}_{d}$, can be defined as the difference between the group and the phase speeds
\begin{equation}
    {\bm{f}_{d}(\bm{k})}\equiv{\bm{v}}_\mathrm{g}(\bm{k})-{\bm{c}(\bm{k})} \,.
    \label{eq:dispersion_factor}
\end{equation}
In general, for a linear dispersion relation, the group velocity $\bm{v}_\mathrm{g}$ coincides with the phase velocity $\bm{c}$ and a wave envelope with the central wavenumber $\bm{k}$ is said to be non-dispersive. In this case, $\bm{f}_{d}= \bm{0}$. For quadratic dispersion relations instead,  $\bm{f}_d(\bm{k})$ represents exactly the directional spreading rate of the wave envelope. In the general case of higher order nonlinear dispersion relations, the wave packet undergoes both a spatial spreading and a distortion, and $|\bm{f}_d|$ is an increasing function of these processes.
In the following, the sign of $\bm{f}_{d}$ will be kept in order to retain information about whether the packet travels faster ($\bm{f}_{d}>\bm{0}$) or slower ($\bm{f}_{d}<\bm{0}$) with respect to the central wavenumber.

\subsection{\label{subsec3:disp_maps}Dispersion maps}

\begin{figure}[htbp!]
	\includegraphics[width=0.82\columnwidth]{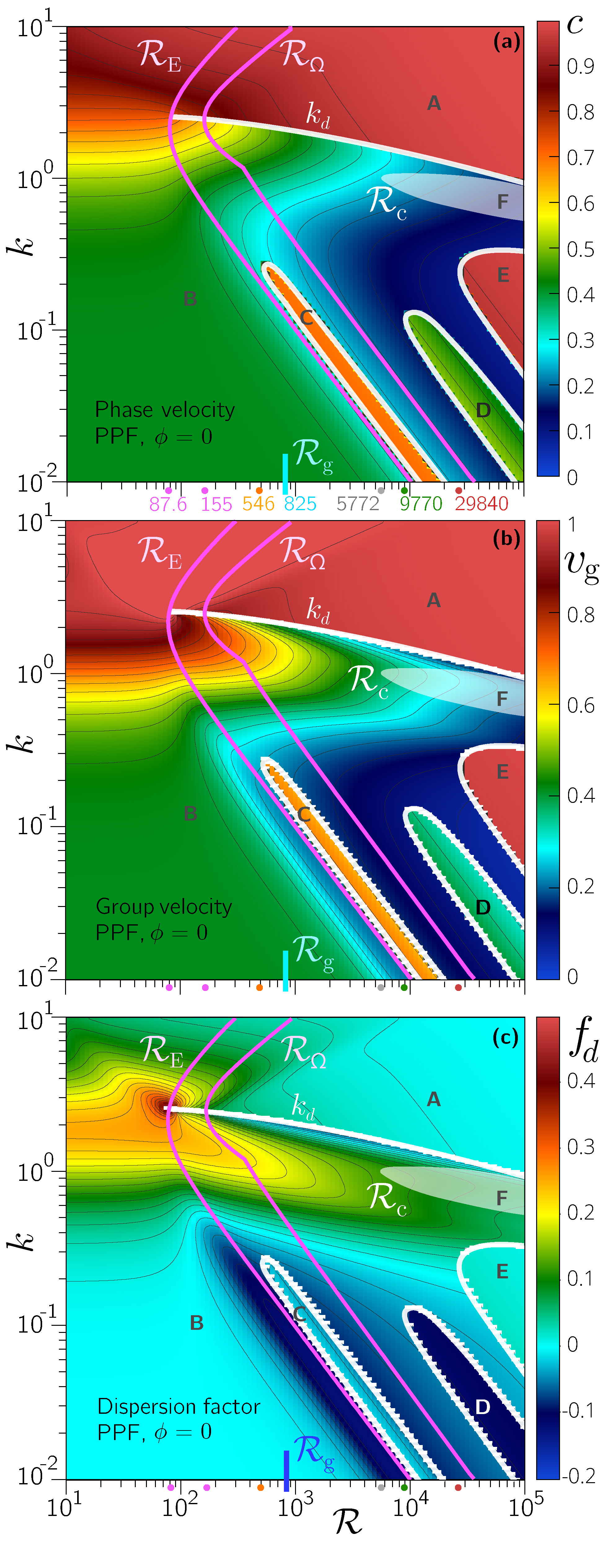}\vskip-13pt
	\caption{Dispersion relation of the least-damped mode of longitudinal waves in PPF, $\mathcal{R}\in[10,10^5]$ and $k\in[10^{-2}, 10]$. The maps contain 100$\times$240 ($\mathcal{R},k$) simulations, uniformly distributed in the log-log space. {(a)} Phase velocity, $\bm{c}$.	{(b)} Group velocity, $\bm{v}_\mathrm{g}$.  {(c)} Dispersion factor,  $\bm{f_{d}=v_\mathrm{g}-c}$. 	$k_{d}$ is a dispersion threshold  \cite{desanti2016} that bounds the non-dispersive \textit{sub-region} \textbf{A}. The pink curves $\mathcal{R}_\mathrm{E}(k), \mathcal{R}_\Omega(k)$, represent the lower bounds for kinetic energy and enstrophy transient growth, respectively \cite{fraternale2018}. In the low-$k$ part of the maps, three  sub-regions \textbf{C}, \textbf{D}, \textbf{E},  have different dispersion properties than the surroundings (\textbf{B}). \textit{Sub-region} \textbf{F} represents the asymptotic instability region.	\label{fig:dispersion_maps}}
\end{figure}

We focus on the dispersion relation of least-damped OS mode, i.e. the mode that experiences the largest exponential growth. These results will be used in Sec. \ref{sec4:wp_asymp_repres} to model the asymptotic dynamics of wave packets in the PPF. Results are presented in Fig. \ref{fig:dispersion_maps}.
The three panels in Fig. \ref{fig:dispersion_maps} show the magnitude of the phase velocity $c$ (a), the group velocity $v_\mathrm{g}$ (b) and the dispersion factor $f_{d}$ (c), respectively, for longitudinal waves ($\phi=0$) seen in their long-term evolution.  
In the maps shown in Fig. \ref{fig:dispersion_maps}, the longitudinal wavenumber ($k\equiv\alpha$) and the Reynolds number are uniformly distributed in the log-space, over a grid of $100\times240$ points respectively, with $\mathcal{R}\in[10, 10^5]$ and $k\in[10^{-2}, 10]$.

The existence of several sub-regions is highlighted in Fig. \ref{fig:dispersion_maps} (labeled with letters \textbf{A} to \textbf{F}). The net separation between such regions is due to the change of identity of the leading OS mode. Note that the boundary of each sub-region has been enhanced for the clarity of visualization. In panel (a), the smallest Reynolds number for which a sub-region is found is also reported.   

A first observation is the sharp separation between the fast non-dispersive waves of sub-region \textbf{A} (${c}\approx{v_\mathrm{g}}\approx1$ for ${k}>{k}_{d}$) and the slow dispersive waves of sub-region \textbf{B} (${k}<{k}_{d}$), which ends up near the nose at $\mathcal{R}_\mathrm{E}(k)\approx87.6$ and $\alpha\approx2.5$. We name the threshold wavenumber $k_d(\mathcal{R})$ (white curve). As mentioned in Sec. \ref{sec1:intro}, the existence of sub-region \textbf{A} was first shown in our previous studies \cite{desanti2016,fraternale2018}.
Interestingly, the map in the neighborhood of  $\mathcal{R}\approx87.6$ and $\alpha=2.5$ is characterized by a high level of dispersion, as can be deduced from $f_d$ in panel (c).
The upper boundaries of the two-dimensional {monotonic stability} regions for the kinetic energy and the enstrophy of perturbation waves are reported in Fig. \ref{fig:dispersion_maps} with the pink curves $\mathcal{R}_\mathrm{E}(k)$ and  $\mathcal{R}_\mathrm{\Omega}(k)$, respectively. The meaning of these thresholds is that transient growths of energy/enstrophy are only possible for larger $\mathcal{R}$ than the respective limit value, whatever initial condition is specified in the initial-value problem \cite{schmid2007,manneville2016}. The analytical derivation of the enstrophy curve can be found in \citet{fraternale2018}.
The white shadowed region in Fig. \ref{fig:dispersion_maps} denotes the well-known unconditional instability region, \textbf{F}. Here, the dominant OS eigenvalue is characterized by $\sigma_i>0$. This region is located in the dispersive part of the graph, just below the curve $k_{d}(\mathcal{R})$, for $\mathcal{R}>5772.22$ \cite{orszag1971}.

A remarkable finding is the existence of the three sub-regions \textbf{C}, \textbf{D}, and \textbf{E}, which have different dispersion features from their surroundings and are located in the dispersive, lower part of the map at $\mathcal{R}>576$. These regions appear tilted at 45$^\circ$ in the log-log plane. It should be noted that the $f=f[\log(k \mathcal{R})]$ trend is recurrent for low wavenumbers in all type of stability maps, not only as regards the monotonic boundaries mentioned previously and the asymptotic phase velocity, but also pertaining to the maximal transient growth of kinetic energy and enstrophy [see also Fig. \ref{fig:maxGZT} in App. \ref{app:A}].
Thus, looking at the map from small to large wavenumbers under fixed flow conditions, that is at constant $\mathcal{R}$, least-damped modes can be observed across a number of fluctuations of the dispersion and propagation properties. As a consequence, waves with distant wavelengths can have the same group speed value producing a linear \textit{dispersive focusing}, which may significantly change as $\mathcal{R}$ varies.

In details, sub-region \textbf{C} is found at $\mathcal{R}>545$ and $k<0.28$. It contains anti-symmetric wall-modes (A-family, but close to S-family modes, according to the classification made by \citet{mack1976}) having intermediate phase speed $c\approx0.7$, pretty much equal to the group velocity $v_\mathrm{g}$. Then, the dispersion level in \textbf{C} is low, and lower than that of the surrounding part of the map. 
Sub-region \textbf{D} is met at $\mathcal{R}>9770$ and $k<0.13$, thus it lies below the unconditional instability region \textbf{F}. Here, the identity of the eigenmodes is the same as in sub-region \textbf{C}. However, these solutions travel with a smaller phase speed, $c\approx0.45$, which slightly depends on the wavenumber $k$,  leading to high dispersion for wave packets containing this range of wavenumbers. In particular, the group speed ($v_\mathrm{g}\approx0.35$) is higher compared to the background region [see Fig. \ref{fig:dispersion_maps}(b)] and smaller than the phase speed $c$ [see panel (c)].
The third sub-region (\textbf{E}) is met at  $\mathcal{R}>29840$ and $k<0.35$. In this case, the least-damped mode belongs to the right branch of the spectrum (P-family), it is an anti-symmetric, fast ($c\approx 1$), and central mode. Therefore, the behavior in this region is quite the same as in the non-dispersive part of the map at $k>k_{d}$ (sub-region \textbf{A}).
A common feature to all the three sub-regions is that they are the only parts of the map where the leading mode (stable, in all cases) is anti-symmetric.
In the case of oblique waves ($\phi\ne0$), we recall that the dispersion relation of the least-damped mode can be deduced from the longitudinal case described in Fig. \ref{fig:dispersion_maps} via the Squire's transformation\cite{squire1933}. 
\begin{equation}\label{eq:squire_transformation}
\omega_\mathrm{3D}(k,\mathcal{R})=\omega_\mathrm{2D}\left(k,\alpha/k\mathcal{R}\right) \,.
\end{equation}
Thus, with increasing $\phi$, the map structure remains unchanged, apart a shift towards higher $\mathcal{R}$, in the log-log plot (this of course does not apply to the curves $\mathcal{R}_\mathrm{E}, \mathcal{R}_\Omega$). Concurrently, both the phase velocity and the group velocity decrease by a factor equal to $\cos(\phi)$, and eventually the phase speed of an oblique wave with wavenumber $\bm{k}$ reads:%
\begin{equation}\label{eq:cosinelaw}
\bm{c}_\text{3D}(\bm{k};\mathcal{R})=\bm{c}_\text{2D}\left(\vert \bm{k} \vert;\mathcal{R}\cos\phi\right)\cos\phi \,.
\end{equation}

\section{\label{sec4:wp_asymp_repres}Wave packet asymptotic representation}

The dispersion relation $\omega(\bm{k}; \mathcal{R})$ described in Sec. \ref{sec3:disp_maps} can be used to obtain an approximate asymptotic representation of a local impulsive disturbance. A wave packet can be formally represented by means of the following Fourier integral
\begin{equation}\label{eq:fourierintegralmt}
\widetilde{q}(\bm{x},t)=\int_{-\infty}^{+\infty}\hat q(\bm{k}) e^{i \theta(\bm{k};\bm{x},t)}\ud\bm{k} \, ,
\end{equation}
where $\widetilde q$ is any relevant physical quantity and $\theta(\bm{k};\bm{x},t)\equiv \bm{k}\cdot\bm{x}-\omega(\bm{k}) t$ represents the phase. Let us consider the Taylor series expansion of the dispersion relation around an arbitrary wavenumber $\bm{k}_0$,
\begin{eqnarray}\label{eq:taylor_omegamt}
\omega(\bm{k})& & =\omega(\bm{k}_0)+\boldsymbol{\nabla}_k \omega\Big|_{\bm{k}_0}\cdot(\bm{k}-\bm{k}_0) \nonumber\\  & &+(\bm{k}-\bm{k}_0)^{T}\bm{H}_\omega\Big|_{\bm{k}_0}(\bm{k}-\bm{k}_0)+\mathcal{O}(\Vert \bm{k}-\bm{k}_0\Vert^3)\,,
\end{eqnarray}
where $(\bm{H}_\omega)_{ij}\equiv\frac{\partial^2 \omega}{\partial k_i \partial k_j}$ is the Hessian matrix. By truncating the series at the linear term, Eq. \ref{eq:fourierintegralmt} becomes
\begin{equation}\label{eq:approx}
\widetilde{q}(\bm{x},t) \approx e^{i\bm{k}_0\cdot\bm{x}-i\omega(\bm{k}_0) t}\int_{-\infty}^{+\infty}\hat q(\bm{k}) e^{i(\bm{k}-\bm{k}_0)\cdot(\bm{x}-\boldsymbol{\nabla}_k \omega|_{\bm{k}_0}t)}\ud\bm{k}\,,
\end{equation}
where the first factor indicates a monochromatic wave developing in time, according to $\exp{[\omega_i(\bm{k}_0) t]}$, and moving with a phase speed $\bm{c}_0=\omega_r(\bm{k}_0)/\Vert \bm{k}_0 \Vert \, \bm{e_k}$, while the second factor represents the wave packet envelope near $\bm{k}_0$, that travels with the group speed $\bm{v}_\mathrm{g}=\bm{\nabla}_k \omega_r|_{\bm{k}_0}$ and dissipates energy in the fluid medium depending on $\bm{\nabla}_k \omega_i|_{\bm{k}_0}$, as recently discussed by \citet{gerasik2010}. Since the viscosity plays a dual role in our system, being the cause of both the flow instability and the damping, it would be interesting to extend to our context the analysis of Ref. \onlinecite{gerasik2010}, including
both the dissipation out-flux and the molecular diffusive process in the interpretation of the complex group speed. The interpretation by \citet{muschietti1993}, where $\bm{\nabla}_k \omega_i$ is related to the differential damping among the Fourier components and thus to the slow drift of the central wave number along the packet trajectory, is also very important in our view and can be considered complementary to that of Ref. \onlinecite{gerasik2010}.
Eq. (\ref{eq:approx}) helps to understand the concept of energy-carrier wave and the role of the group velocity, but it is no longer useful to describe the dispersion phenomenology which can lead to a distortion of the initial envelope shape. Looking for the asymptotic form (large $t$ with $\bm{x}/t$ of order unity),  the \textit{steepest-descent method} (also referred to as the \textit{saddle-point method}) \cite{duistermaat1974,bender1978book} leads to the following expression 
\begin{eqnarray}\label{eq:stationaryphasemt}
\widetilde{q}(\bm{x},t)&\approx& \hat q(\boldsymbol{\kappa}) \Big( \frac{2\pi}{t}\Big)^\frac{n}{2}\sqrt{\frac{1}{|{\det}(\bm{H}_\omega(\boldsymbol{\kappa}))|}} \nonumber \\ & & \times e^{i \left\{\boldsymbol{\kappa}\cdot\bm{x}-\omega(\boldsymbol{\kappa}) t-\frac{\pi}{4}{\mathrm{sign}} [{\det}(\bm{H}_\omega)]\right\} } \nonumber \\ & &=A(\bm{x},t)e^{i\chi(\bm{x},t)} \, ,\label{eq:stationaryphase2}
\end{eqnarray}
where $n$ is the space dimension ($n=2$ in the case of planar waves) and  $\boldsymbol{\kappa}$ is the specific wavenumber which makes the phase $\theta$ of the integral (\ref{eq:fourierintegralmt}) stationary:
\begin{align}\label{eq:rayscomplex_m}
\boldsymbol{\kappa}:\hskip10pt&\boldsymbol{\nabla}_k\theta(\boldsymbol{\kappa})\equiv\bm{x}-\boldsymbol{\nabla}_k\omega(\boldsymbol{\kappa})t=\bm{0} \\ &\Rightarrow \hskip5pt
\begin{cases}\label{eq:rayscomplex_cond}
\bm{x}/t=\boldsymbol{\nabla}_k \omega_{r}(\boldsymbol{\kappa})\equiv\bm{v}_\mathrm{g}\\
\bm{0}=\boldsymbol{\nabla}_k \omega_{i}(\boldsymbol{\kappa})\,.
\end{cases} 
\end{align}
The $\tfrac{1}{2}$ exponent at the second and third factors in Eq. (\ref{eq:stationaryphasemt}) is related to the first nonzero term in the expansion (\ref{eq:taylor_omegamt}), which here is assumed to be the term including second derivatives. If Eq. (\ref{eq:rayscomplex_m}) has solutions $\boldsymbol{\kappa}(\bm{x},t)$, this means that a specific wavenumber, $\boldsymbol{\kappa}$, dominates the packet  in the physical space defined by Eq. (\ref{eq:rayscomplex_cond}). In particular, the disturbance is a  space- and time-dependent wave packet, whose local frequency is $\omega_r(\bm{\kappa})$ at the space location $\bm{x}$ and time instant $t$. In the general case of complex-valued dispersion relation, the disturbance experiences  a transient growth or damping, which is given by the factor $\exp{[\omega_i(\bm{\kappa}) t]}$, in addition to the amplitude factor $A(\bm{x},t)$ of Eq. \ref{eq:stationaryphase2}.
Note that early applications of this method to the field of hydrodynamics for shear flows are found, e.g., in Refs. \onlinecite{benjamin1961,criminale1962,gaster1965,craik1981}. Indeed, for homogeneous media, the theory of wave groups with slowly varying properties leads to the conservation of wavenumbers and wave angles along straight lines in the $\bm{x}-t$ plane. Differently than in the previous studies, we will not use analytic models of the dispersion relation. Instead, we use the exact  results shown in Sec. \ref{sec3:disp_maps}, as described below.

\begin{figure*}
	\includegraphics[width=\textwidth]{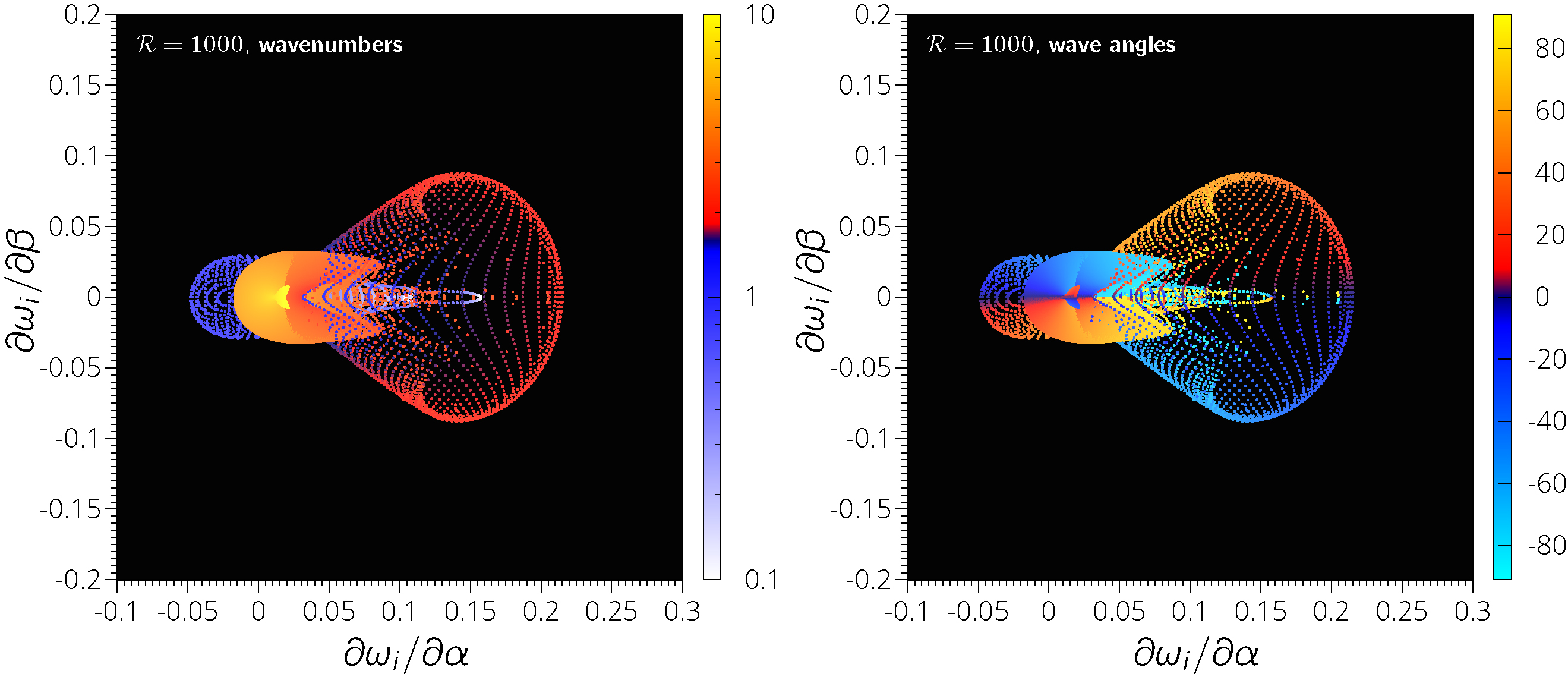}
	\caption{Gradient of the imaginary part of the dispersion relation for the least damped OS mode for PPF at  $\mathcal{R}=1000$. Visualization in terms of wavenumber (left panel) and  wave angle (right panel). }
	\label{fig:VGimag}	
\end{figure*}

\begin{figure*}
	\includegraphics[width=0.8\textwidth]{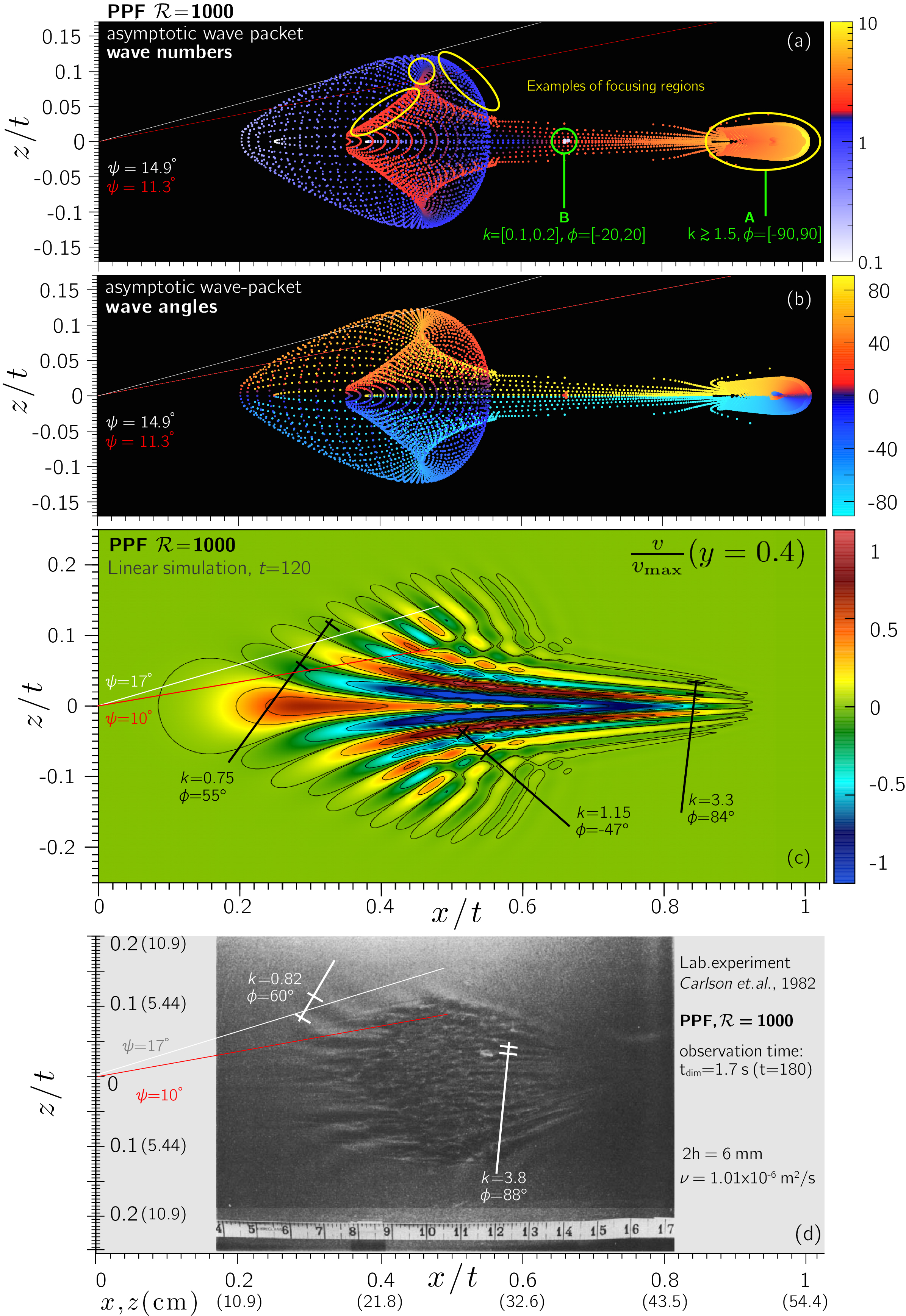}
	\caption{Wave packet long-term propagation scheme,  numerical simulation and laboratory visualization at $\mathcal{R}=1000$. (a) Wavenumber propagation. The regions of most intense wave focusing are highlighted with yellow and green circles. (b) Wave angle propagation. (c)  Numerical simulation of the linearized initial-value problem [Eqs. (\ref{eq:wall_normal_vel}-\ref{eq:wall_normal_vort}), see also App. \ref{app:A}] for an initial Gaussian disturbance. Two visualization movies are available (Multimedia view).  (d) Laboratory visualization adapted from Fig. 5 of \citet{carlson1982}. Here, the axes have been normalized with $t_\mathrm{dim}=1.7~s$ derived from information in Fig. 11 of the same study, and rescaled to match the other panels.}
	\label{fig:main_figure}	
\end{figure*}

Results from the implementation of the saddle-point method in PPF are shown in Figs. \ref{fig:VGimag} and \ref{fig:main_figure}(a,b) (Multimedia view) for $\mathcal{R}=1000$, and in Fig. \ref{fig:medusa_lin_kappa_AllRe} for $\mathcal{R}=500, 1000, 2000,$ and 4000. As shown in Sec. \ref{sec3:disp_maps}, the group velocity of the least-damped wall-normal velocity mode is computed from the exact dispersion relation of 3-D waves. These figures show the long-term space distribution, in the $x-z$ plane, of a wave packet initially localized at the origin ($x=z=0$). The coordinates of Figs. \ref{fig:main_figure} and \ref{fig:medusa_lin_kappa_AllRe}  are normalized over time, so that they represent the group velocity. Therefore, the spreading rates can be directly inferred from the figures. For each $\mathcal{R}$, the wave packet is made of a discrete number of vector wavenumbers, $N_w=57600$. In particular, we set a uniform grid for both the streamwise and the spanwise wavenumber components, $\alpha\in[10^{-2}, 10],\ \beta\in[-5,5]$, and the grid spacing is $\Delta \alpha=\Delta \beta=0.0417$. For each vector wavenumber, the least stable OS mode is computed, and then propagated according to Eq. (\ref{eq:rayscomplex_cond}). Therefore, each point in our figures represents an individual wave component. The corresponding wavenumber and wave angle are color-coded.

Figure \ref{fig:VGimag} shows the imaginary part of the group velocity. Due to the the limited differential damping among the OS leading modes, the normalized components of the imaginary part of the group velocity are small. For instance, at $\mathcal{R}=1000$ the $x$-component is in the range [-0.03,0.08] for about 95\% of the wavenumbers considered in this study, while the $z$-component is within [-0.03,0.03] [see Fig. \ref{fig:VGimag}]. Therefore, since $\boldsymbol{\nabla}_k \omega_r(\boldsymbol{\kappa})/\boldsymbol{\nabla}_k \omega_i(\boldsymbol{\kappa})\ll 1$, the second condition in Eq. (\ref{eq:rayscomplex_cond}) is found to be approximately satisfied by most of wavenumbers.

Figure \ref{fig:main_figure} (Multimedia view) shows the asymptotic shape of the wave packet and compares it to both a numerical simulation of the linearized initial value problem and to a laboratory experiment. Specifically, panel (a) in Fig. \ref{fig:main_figure} shows the results from Eq. (\ref{eq:rayscomplex_cond}) in terms of the wavenumber magnitude, while panel (b) displays the wave angle. The unsteady linear evolution of a 3-D localized wave packet in PPF is investigated by integrating the linearized Navier-Stokes equations [Eqs. (\ref{eq:wall_normal_vel}-\ref{eq:wall_normal_vort})] via our semi-analytic code based of Fourier and Chandrasekhar functions [see App. \ref{app:A} for further details]. We should mention that for an extension of our analysis to weakly non-parallel flows, alternate stability analysis methods should be considered, such as the Parabolized Stability Equations (PSE) or the Wentzel-Kramers-Brillouin-Jeffreys (WBKJ), or (bi- and tri-)global stability methods in general \citep[e.g., see ][]{petz2011,bistrian2013,nastro2020}.

Results of our numerical simulation at $\mathcal{R}=1000$ are shown in panel (c). Here, we choose as initial condition a localized (at $x=z=0$) wall-normal velocity disturbance with a Gaussian distribution in the $x-z$ plane. The initial distribution along the $y$ axis is such to guarantee a large energy growth rate. Technical details concerning simulation parameters are reported in App. \ref{app:A}. In panel (d), we reproduce Fig. 5 of the well known laboratory experiment carried out by \citet{carlson1982} in 1982, showing a developed spot where turbulence and waves coexist. Note that this visualization has been accurately scaled in order to allow comparison with our simulation. To achieve this, information from Fig. 11 in Ref. \onlinecite{carlson1982} has also been used. 
In terms of normalized time, the experiment visualization is at $t=180$. The choice of $t=120$ for the simulation of panel (c) is essentially related to the fact that it is an intermediate state [intermediate asymptotics, see 1996 Barenblatt monography\cite{barenblatt1996}] between the early transient - the short period when the morphology changes very rapidly - and the long term, when the linear packet has decayed. This time may depend on the flow parameters and initial conditions.  At $t=120$ the packet's propagation properties are nearly asymptotic, as shown in panels (a,d) of Fig. \ref{fig:spreading_rates}, which allows the comparison with the asymptotic representation. Note also that, the maximal energy and enstrophy gain is reached at $t\approx 20$ ($t_\mathrm{dim} \approx 0.2~s$), as shown in panels (e,f), of Fig. \ref{fig:spreading_rates}, so that the linear solution decays at large times. The comparison with experiments is meaningful only in the early/intermediate stage of these structures. Although in the experiment the spot at $t\approx120$ has already developed the turbulent core, it is interesting to look at the laminar wave regions surrounding the core.
\begin{figure*}
	\includegraphics[width=0.8\textwidth]{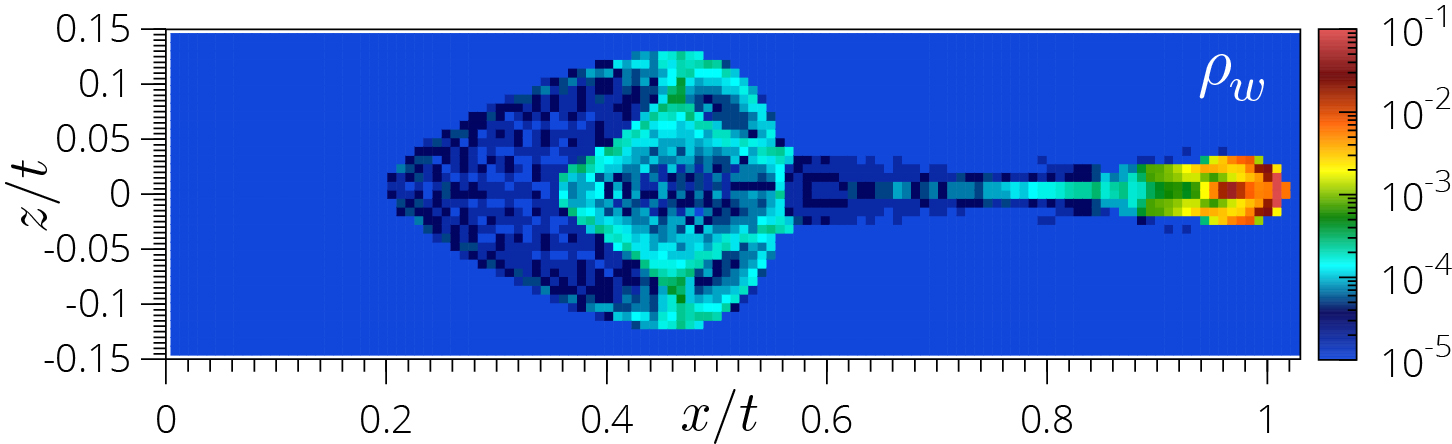}\vskip-5pt
	\caption{Wave focusing, shown via the normalized histogram of wave components, see Eq. (\ref{eq:density_fun}), for the wave packet asymptotic representation of Fig. \ref{fig:main_figure}(a,b), $\mathcal{R}=1000$. }
	\label{fig:wdens}	
\end{figure*}
A first observation is that most of the typical features of the 3-D wave packet are recovered by the representation of panels (a,b): the arrow pattern of the spot is made up of a slower almond-shaped rear part, a faster front and a streaky tongue connecting these two regions. We highlight that a remarkable quantitative agreement is achieved between dominant wavenumbers and wave angles obtained from the asymptotic representation and those observed in the laminar wave patches of real spots. Typical vector wavenumbers for both the simulation and the laboratory experiment are reported in panels (c) and (d) of Fig. \ref{fig:main_figure} (Multimedia view).
The bulb-shaped front is difficult to observe in a laboratory because its wave components are highly damped. However, the nature of these nondispersive modes,  makes them relevant from the enstrophy viewpoint, as we have recently shown \cite{fraternale2018} [see also Ref. \cite{duguet2010b}, in the context of pipe flows]. The core of the wave packet can be recognized as the rhomboid-shaped region at $0.35\lesssim x/t\lesssim0.5$. Note that in panel (a) the color bar is set so that the non-dispersive components $k>k_{d}\approx 2$ and the inner rhomboid part are highlighted. This pattern is typical of both the early- and the self -sustained evolution of the spot's life [see, for instance, Fig. 4 in Refs. \onlinecite{lemoult2013} and \onlinecite{lemoult2014}, Fig. 15 in Ref. \onlinecite{klotz2017a}]. 

Wave focusing can be measured by evaluating the concentration of wave components at specific locations in the $x-z$ plane of the physical space. In practice, we discretize the $x-z$ plane in $64\times32$ bins and compute the normalized histogram:
\begin{equation}
    \rho_w(x_j,z_j) \equiv \frac{n(x_j,z_j)}{N_w} \,,
    \label{eq:density_fun}
\end{equation}
where $n(x_j,z_j)$ represents the number of waves in the $j$\textsuperscript{th} bin, centered at $(x_j,z_j)$, and $N_w=57600$ denotes the total number of waves used in the asymptotic representation. Therefore, $\rho_w$ represents the probability of finding -- at a specific spatial location and in the long term -- a wave component from the initial packet. Results are shown in Fig. \ref{fig:wdens}, for the same parameters used in Fig. \ref{fig:main_figure}(c). It should be reminded that $\rho_w$  is derived from the asymptotic model, in the modal-stability framework. Therefore, it represents the probability of finding -- at a specific space location -- a wave component from the initial packet in the long-term limit. As discussed previously, it approximately represents the wave distribution in the intermediate transient. In the very short early stage its distribution can be different and rapidly changing from the initial state where all waves are focused in the origin. It would be interesting to investigate the time dependence of $\rho_w$. This requires the computation of time-dependent dispersion relations, in the framework of the non-modal stability analysis. This task is delicate, because in the early transient the phase speeds may experience  rapid fluctuations \citep{desanti2016}, and will be postponed to a future study.  Interestingly, we observe a large wave concentration in the front, and at the wingtips of the inner rhomboid-shaped region. 

At the wingtips, the wave components are quite oblique with angles around $75^\circ$. These waves are the most algebraically unstable, that is, they experience the largest growth in transient kinetic energy and enstrophy in the intermediate term. Moreover, as shown here, wave focusing is significant at the wingtips. This is consistent with the experimental observation that the transition to turbulence is first triggered at the wingtips \citep{carlson1982}. Waves at the spot's leading edge are  more oblique, nearly orthogonal to the basic flow ($\phi\approx85^\circ$). 

The green circle in panel (a) of Fig. \ref{fig:main_figure} highlights the focusing between short non-dispersive wave components with $k\approx 2$ and a few long, non-dispersive, waves belonging to region \textbf{C} of the dispersion maps of Fig. \ref{fig:dispersion_maps}. These long waves appear as white dots concentrated at the centerline where $x/t\approx0.7$. Together with the highly-focused short waves of the wave packet front, these long modes may play a role in the generation of the intense and long-lasting shear layer that connects the front of the wave packet to its core, the so-called ``permanent scar'' observed by Landhal \cite{landahl1975,landahl1980}. Other focusing loci are indicated with yellow circles in the same panel [see Fig. \ref{fig:main_figure}(a)].

The asymptotic spreading rates can be directly inferred from the propagation scheme shown in Fig. \ref{fig:main_figure}(a,b). The lines that connect the origin to the point of maximum spreading (white line) for the entire packet and (red line) for the inner rhomboid-shaped region can be observed. These lines define the spreading half-angle $\psi_\mathrm{spread}$. Figure \ref{fig:medusa_lin_kappa_AllRe} shows the Reynolds number effect on the asymptotic shape of wave packets, such as that presented in Fig. \ref{fig:main_figure}(a,b). 
\begin{figure*}
   \includegraphics[width=\textwidth]{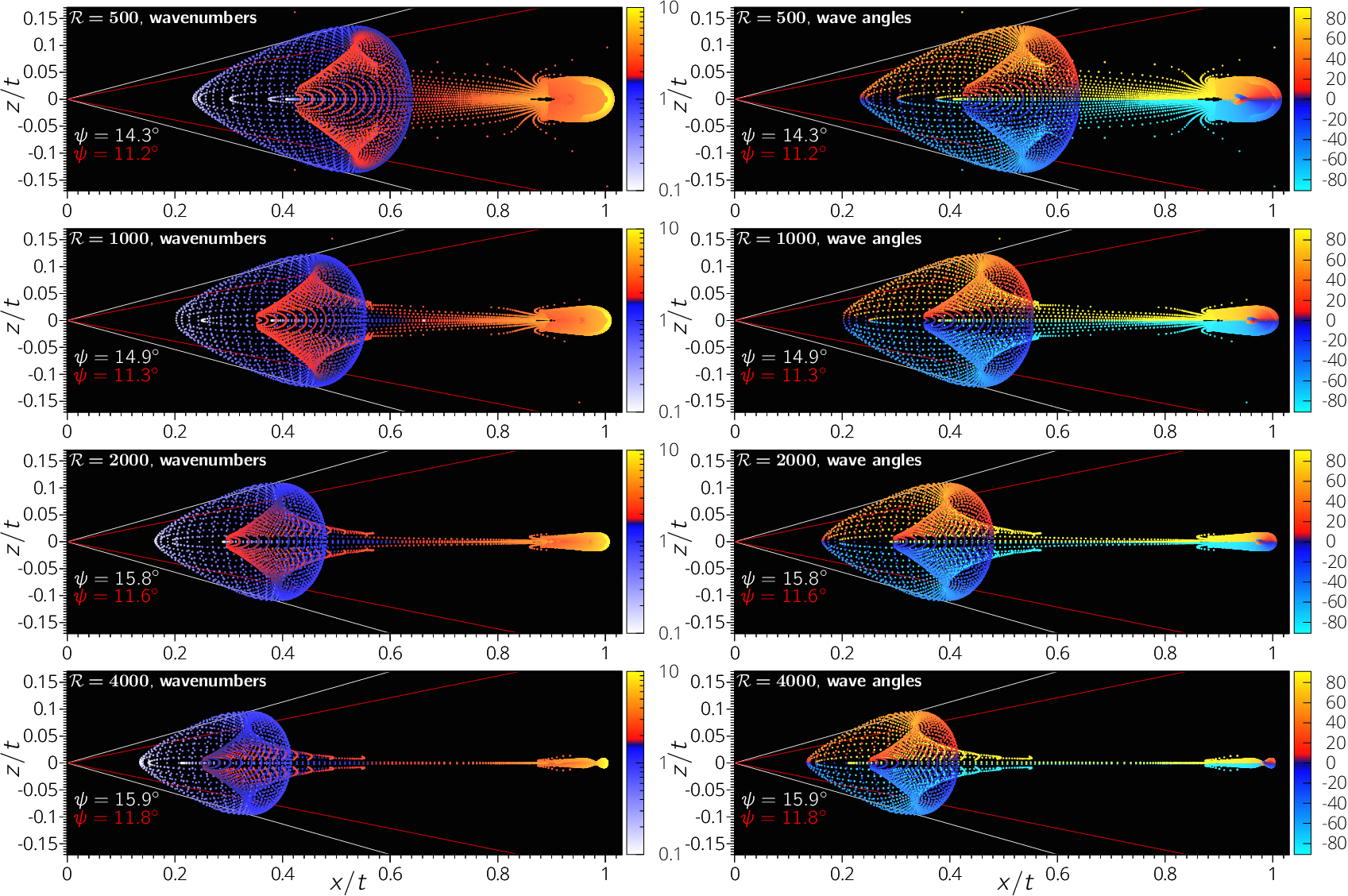}
    \caption{Asymptotic wave packet representation: Reynolds number effect. The propagation of the wavenumber magnitude (left) and the wave angle (right) are shown for $\mathcal{R}=[500,1000,2000,4000]$ (top to bottom).}
    \label{fig:medusa_lin_kappa_AllRe}
\end{figure*}
As $\mathcal{R}$ increases, an increment of the spreading angle $\psi_{\mathrm{spread}}$ is found to be related to the concomitant narrowing and slowing of the wave packet core as $\mathcal{R}$ increases, which results in an elongation of the arrow-shaped structure. Calculated values of $\psi_\mathrm{spread}$ agree well with laboratory observations, as evidenced by the comparison with Fig. \ref{fig:main_figure}(d). Streamwise spreading in the simulation is computed by searching for the portion of domain that contains $95\%$ of the energy. The boundaries of this region are named
${x}_{\mathrm{rear}}$ and ${x}_{\mathrm{front}}$, respectively. In particular, for a generic quantity $\widetilde q$, they are defined as the location of the $2.5$\textsuperscript{th} and $97.5$\textsuperscript{th} percentiles of the $xy$-reduced squared variable:
\begin{equation}
    {x}_{\mathrm{rear}, \mathrm{front}}(t):\hskip4pt \int_{-\infty}^{x_{\%}}\langle \widetilde{q}^2\rangle_{z} \ud x= C \int_{-\infty}^{\infty} \langle\widetilde q^2\rangle_{z} \ud x \,,
\end{equation}
where $\langle \widetilde q \rangle_{z}(t,x,y)={L_{z}}^{-1}\int \widetilde q\ud z$ ($L_z$ is the domain size in the spanwise direction), $C = 0.25$ and  $0.975$ for the rear and front, respectively. 
Analogous metrics are defined in details in App. \ref{app:A}, and a detailed review of these specific spatial features corresponding to our simulations [see Fig. \ref{fig:main_figure}(c) (Multimedia view)] is reported in Tab. \ref{tab:WPspreading500-1000} and in Fig. \ref{fig:spreading_rates} of App. \ref{app:A}. A review of literature experiments is given in Tab. \ref{tab:WPspreading_literature}. Here, the values are either explicitly stated by the authors, or extracted from visualizations. Most of them refer to turbulent/laminar interfaces. It can be observed that the rear of the inner turbulent region corresponds to the center of our linear spots, and that it moves at a speed nearly equal to one half of the centerline channel speed. The spot front moves faster, with a speed about 0.7-0.8$U_{C}$, that is almost equal to the values found for ${x}_\mathrm{front}$ in the linear spots. The lateral spreading rates we observe compare well with the values obtained from experiments that lie in the range [0.06, 0.12], while  ${z}_\mathrm{spread}$ is somewhere between 0.06 and 0.12 depending on which field component considered. If the kinetic energy or the enstrophy are considered for this computation, the spreading half-angle is about $10^\circ$. This value is also close to that measured from the turbulent core of laboratory spots. 
\begin{table*}
\begin{ruledtabular}
\begin{tabular}{lcccccccc} 
		\multicolumn{9}{c}{\hskip0pt {\bf $\mathcal{R}=1000$}} \\ \hline
		&  ${\widetilde u}$  &${\widetilde v}$ &${ \widetilde w}$&${ \langle E\rangle_y}$&$ {\widetilde \omega}_{{x}}$ &$ {\widetilde \omega}_{{y}}$&$ {\widetilde \omega}_{{z}}$&$ {\langle\Omega \rangle}_{{y}}$\\ 	\hline
		${{x}}_{\mathrm{G}}$                   &     0.51       &   0.45   & 0.50  &0.54& 0.50 &0.57 &0.56& 0.56\\ 
		${{x}_\mathrm{rear}}$             &    0.31      &   0.24      & 0.38 & 0.33&0.38&0.42 &0.39 &0.35 \\ 
		${{x}_\mathrm{front}}$           &    0.74  &    0.65     & 0.79  & 0.80 &  0.82& 0.73&0.79 &0.82\\ 
		${{z}_\mathrm{spread}}$             &    0.08    &    0.12     &  0.09 & 0.04 &0.08  &0.05&0.08&0.05\\ 
		${\psi}_\mathrm{spread}$    &   10.4$^\circ$      &     17.4$^\circ$       &   15.3$^\circ$  & 10.1$^\circ$&14.1$^\circ$ &  8.50$^\circ$& 9.92$^\circ$ & 10.3$^\circ$\\ 
	\hline
	\end{tabular}
		\caption{\label{tab:WPspreading500-1000}
		Spreading rates at $\mathcal{R}=1000$ obtained as the time average for $t\in[15,35]$. The spreading rates of the field components are computed at a fixed distance from the wall, $y=-0.6$, while the kinetic energy and the enstrophy are $y$-averaged.}
\end{ruledtabular}
\end{table*}
\begin{table*}
\begin{ruledtabular}
\begin{tabular}{lcccccccc}
		& ${\mathcal{R}}$&${{x}_{\mathrm{G}}}$  &${{x}_\mathrm{rear}}$&${{x}_\mathrm{front}}$&${{z}_\mathrm{spread}}$&${\psi}{_\mathrm{turb}}$ \\ 	\hline
		C-1982 \cite{carlson1982}                              &1000             & 0.5             &0.33             & 0.6              & -                 &8$^\circ$\\ 
		A-1986 \cite{alavyoon1986}                      & 1100-2200   &-                  &0.62-0.52    & 0.75-0.8      &-                  &6$^\circ$-12$^\circ$\\
		HA-1987 \cite{henningson1987a}              & 1200-3000   & 0.65            & 0.56            & 0.83            &   0.12        &7$^\circ$-15$^\circ$\\ 
		KA-1990 \cite{klingmann1990}                    & 1600             &0.65            &0.55           & 0.7-0.85      &0.09-0.2        &8$^\circ$\\ 
		HK-1991 \cite{henningson1991}                            &   1500            &0.64            &0.55          &  0.70-0.8       & 0.08-0.12    &8$^\circ$-9$^\circ$\\ 
		K-1992 \cite{klingmann1992}                                           & 1600            & 0.65-0.7       &0.55   &0.82         &  0.06                   &6$^\circ$-8$^\circ$\\ 
		LAW-2013 \cite{lemoult2013}                                           &  2000            &   0.66      &  0.54  &     0.84     &     -                 & -\\
		LAW-2013 \cite{lemoult2013}                                           &   3000           &   0.49      &  0.62  &     0.84     &     -                 & -\\ \hline
	\end{tabular}
		\caption[Spreading rates of localized perturbations in the literature on plane Poiseuille flow.]{\label{tab:WPspreading_literature}
		Spreading rates (normalized by the centerline velocity $U_\mathrm{C}$) of localized perturbations and turbulent spots from literature experiments on PPF. These values mostly refer to the turbulent/laminar interface of developed self-sustaining spots. Here, the effect of the Reynolds number $\mathcal{R}$ on the wave packet's shape is shown in Fig. \ref{fig:medusa_lin_kappa_AllRe}.}
\end{ruledtabular}
\end{table*}

\section{\label{sec5:conclusions}Conclusions}

We have shown the existence of linear dispersive wave focusing inside one of the archetypal Navier-Stokes incompressible shear flows, the plane Poiseuille flow (unstratified and in absence of a background rotation). This was done by investigating the dispersion relation of least-damped, small-amplitude perturbation waves in this flow, for a three-decade range of wavenumbers and a four-decade range of Reynolds numbers. 

Looking at the dispersion map in the limit of least-damped waves, we have shown the existence of several sub-regions with remarkably variable levels of dispersion across the parameters space. This particular scenario yields dispersive focusing. We have deduced a propagation scheme, based on the saddle point asymptotic approximation and by using the directional distribution of the numerically computed asymptotic group velocity. This model has proved able to recover the morphology and propagation properties of wave packets under linearized dynamics. We also highlight that several of the propagation features shown here are also observed in laboratory experiments and in direct numerical simulations, where transition to turbulence can occur for Reynolds numbers above the global stability threshold, and laminar flow coexists with turbulent patches. In particular, the analysis suggests a correlation between the rhomboid-shaped region in our model, where focusing is high, and the region where strong nonlinear coupling is more likely to be triggered in the real system. In other dynamical contexts, in particular that of surface water waves, the propagation pairing between long and short waves and the mutual influence of their angle of inclination have led the way to a new understanding into the analysis of both linear and nonlinear wave interactions.

This dynamical aspect has not been considered in great detail in the context of turbulence transition in shear flows. This likely occurred because wavenumbers with values close to the region of the stability map where unconditional instability takes place were predominantly investigated. In the sub-critical transitional context, fully developed turbulence and waves coexist, a scenario which goes beyond the so-called wave-turbulence. The  dispersive focusing in the wave packet's early evolution, shown in this study, leads to the conjecture that this phenomenon  may play an important role in promoting nonlinear wave interaction and in the consequent transition to turbulence, a conjecture that opens up a new investigation perspective for the context of nonlinear dispersive focusing of perturbation waves in shear flows.

\begin{acknowledgments}
We wish to acknowledge the Department of Control and Computer Engineering (DAUIN) of Politecnico di Torino for providing computational resources at \url{HPC@POLITO}. FF acknowledges support from MIUR, under the postdoctoral program FOIFLUT (2015--2018),  grant 37/17/F/AR-B (Politecnico di Torino, DISAT). FF also acknowledges UAH--CSPAR for support during the revision of the paper.  DT acknowledges support from Politecnico di Torino, grant POLITO--DISAT 54\_RBA17TD01.
\end{acknowledgments}

\section*{Data availability}

Data  supporting  the findings of this study are available from the corresponding author upon reasonable request.

\appendix

\section{Numerical simulations} \label{app:A}

\subsection{Methods for the initial-value problem}

Numerical simulations [see Fig. \ref{fig:main_figure}(c) (Multimedia view)] have been performed via a MATLAB software, built to solve the three-dimensional, linearized, incompressible Navier-Stokes equations in planar channel flows. Initial perturbations of arbitrary shape can be specified. The periodic boundary conditions in the $x$ and $z$ directions allow to use a Fourier spectral numerical method.

A pre-processing routine specifies the base flow and the simulation parameters (the Reynolds number $\mathcal{R}$, the domain size $L_{x}$ and $L_{z}$, the time and space discretization and the initial conditions in the physical space). 
In details, we represent the solution for any perturbation variable $\widetilde q$ of a general, 3-D, small perturbation as
\begin{equation}\label{eq:integral_xz}
\widetilde q(\bm{x},t)=\mathrm{Re}\Bigg\{\frac{1}{(2\pi)^2}\iint a_{q}(t,y;\alpha,\beta) e^{i\alpha x+i\beta z} \ud \alpha\ud\beta\Bigg\}\,,
\end{equation}
where $\mathrm{Re}$ stands for the real part. It was convenient to separate variables and split $a_{q}$ in two factors:
\begin{equation}\label{eq:coefficients}
a_{q}(t,y;\alpha,\beta)=\hat q (t,y;\alpha,\beta)\hat f(\alpha,\beta)\,,
\end{equation}
where $\widehat q$ is the complex-valued Fourier coefficient of an individual wave [that is, the solution of the Orr-Sommerfeld/Squire initial-value problem, Eqs. (\ref{eq:wall_normal_vel}-\ref{eq:wall_normal_vort})]. The expression $\widehat f(\alpha,\beta)$ gives the distribution in the $x$-$z$ plane, where we have set a Gaussian distribution of thickness tuned by the parameter $\gamma=0.1$, in order to simulate a localized 3-D impulsive disturbance. Further details on the initial condition chosen for the simulation shown in the present study are reported below in Sec. \ref{sec:initial_condition}. Numerically, the integral (\ref{eq:integral_xz}) is discretized as:
\begin{align}
\widetilde q(x,y,z,t)=\mathrm{Re}\Big\{\frac{1}{N_{x} N_{z}}&\sum_{j=-\frac{N_{x}}{2}}^{\frac{N_{x}}{2}-1} \sum_{k=-\frac{N_{z}}{2}}^{\frac{N_{z}}{2}-1} a_{q}(t,y;\alpha,\beta) e^{i\alpha_j x+i\beta_k z} \Big\}\,,
\end{align}
where $N_x$ and $N_z$ are the number of grid points and $\alpha_j=\frac{2\pi j}{N_{x} \Delta x}$, $\beta_i=\frac{2\pi i}{N_{z} \Delta z}$ are  the discrete streamwise and spanwise wavenumbers, respectively. The temporal evolution of the individual wave components is analytically described, therefore the time grid can be arbitrary and has no effect on the accuracy of the computation. A fine grid is needed for post-processing purposes only, such as movie visualizations or the computation of spreading rates. A summary of the chosen simulation parameters for the linear spot of Fig. \ref{fig:main_figure}(c) is reported in Tab. \ref{tab:WPparameters}.
\begin{table}
	\begin{tabular}{lc} 
		\hspace{1cm}3-D linear \\ 
		\hline
		${x_0}$                 &  0 \\ 
		${z_0}$                 &  0 \\ 
		${\gamma}$          &  0.1  \\
		${\mathcal{R}}$              &  1000  \\
		${L_{x}}$        &  120  \\ 
		${L_{y}}$        &  2  \\ 
		${L_{z}}$        &  60  \\ 
		${N_{x}}$        &  256 \\
		${N_{y}}$        &  129  \\ 
	   ${N_{z}}$         &  128  \\  
	   ${T_\mathrm{max}}$       &  120  \\ 
	   ${\Delta t}$             &  0.3  \\ \hline
	\end{tabular}
	\caption{\label{tab:WPparameters}
		Wave packet parameters for the simulation of Fig. \ref{fig:main_figure}(c) in the main text. $(x_0,z_0)$ is the initial spot's location in $x$-$z$ plane and $\gamma$ is the standard deviation of the Gaussian function used as initial condition to simulate the localized 3-D spot [see Eq. (\ref{eq:Gaussian_function}) for further details]. $L_x \times L_y \times L_z$ is the dimensionless domain size, containing $N_x\times N_y\times N_z$ computational grid points.  $T_\mathrm{max}$ is the dimensionless final time and  $\Delta t$ is the time step.}
\end{table}

The processing core is dedicated to the computation of the temporal evolution of individual Fourier components. It is based on a semi-analytical solution of the Orr-Sommerfeld and Squire initial-value problem via the Gal\"erkin method. Here, an eigenfunction expansion with Chandrasekhar-Reid functions \cite{chandrasekhar61} is used to represent the solution in the wall normal direction $y$. A major issue related to the OS-Squire initial value problem is the intrinsic high sensitivity of the OS operator to numerical perturbations, especially at large $\mathcal{R}$ and wavenumbers \citep{rsh1993}. This may cause large inaccuracy in the computation of the eigenvalues at the intersection of the branches. Whilst the least-damped modes are much less affected by such sensitivity, it can significantly affect the numerical solution during the early/intermediate transient. In fact, it is intrinsically related to the non-orthogonal nature of the eigenvectors. We prevent this issues by using  $300$ modes in the expansion. The full description of the method can be found in the Supplementary Material of Ref. \onlinecite{desanti2016} and in App. A of Ref. \onlinecite{fraternale_phdthesis}. This method has also been used in our recent study on the wave enstrophy \cite{fraternale2018}. Source codes are available  at \url{https://areeweb.polito.it/ricerca/philofluid/software.html}.

At a second stage, a dedicated routine performs the inverse Fourier transform in order to obtain the perturbation velocity and vorticity fields in the physical space. Post-processing routines are used to compute the wave packet center and the spreading rates, and to produce movies. 

\subsection{Initial perturbation for a localized 3-D spot\label{sec:initial_condition}}
\begin{figure*}
   \includegraphics[width=0.7\textwidth]{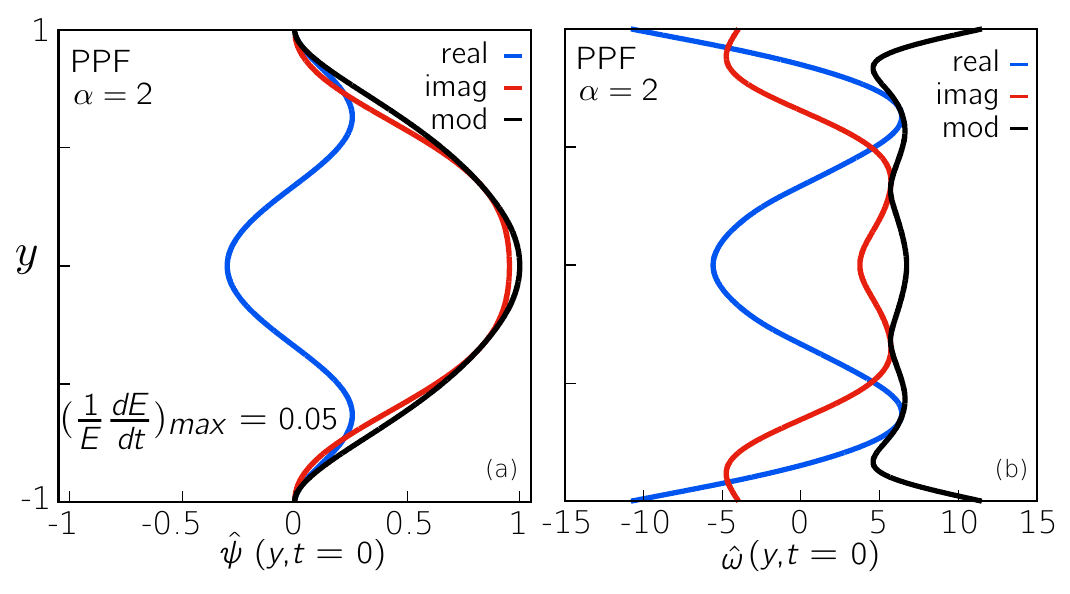}
    \vskip-8pt
    \caption{Initial optimal condition, y-distribution. This figure is part of Fig. 8 in Ref. \onlinecite{fraternale2018}, and it is here reported for the reader's convenience. The optimization procedure seeks the smooth perturbation which maximizes the kinetic energy growth rate  $E^{-1} (\ud E/\ud t)$, at fixed values of $k$ and $\mathcal{R}$, while exciting both symmetric and anti-symmetric Orr-Sommerfeld modes.  Zero initial wall-normal vorticity is set.} 
    \label{fig:initial_v}
\end{figure*}
\begin{figure*}
\includegraphics[width=0.85\textwidth]{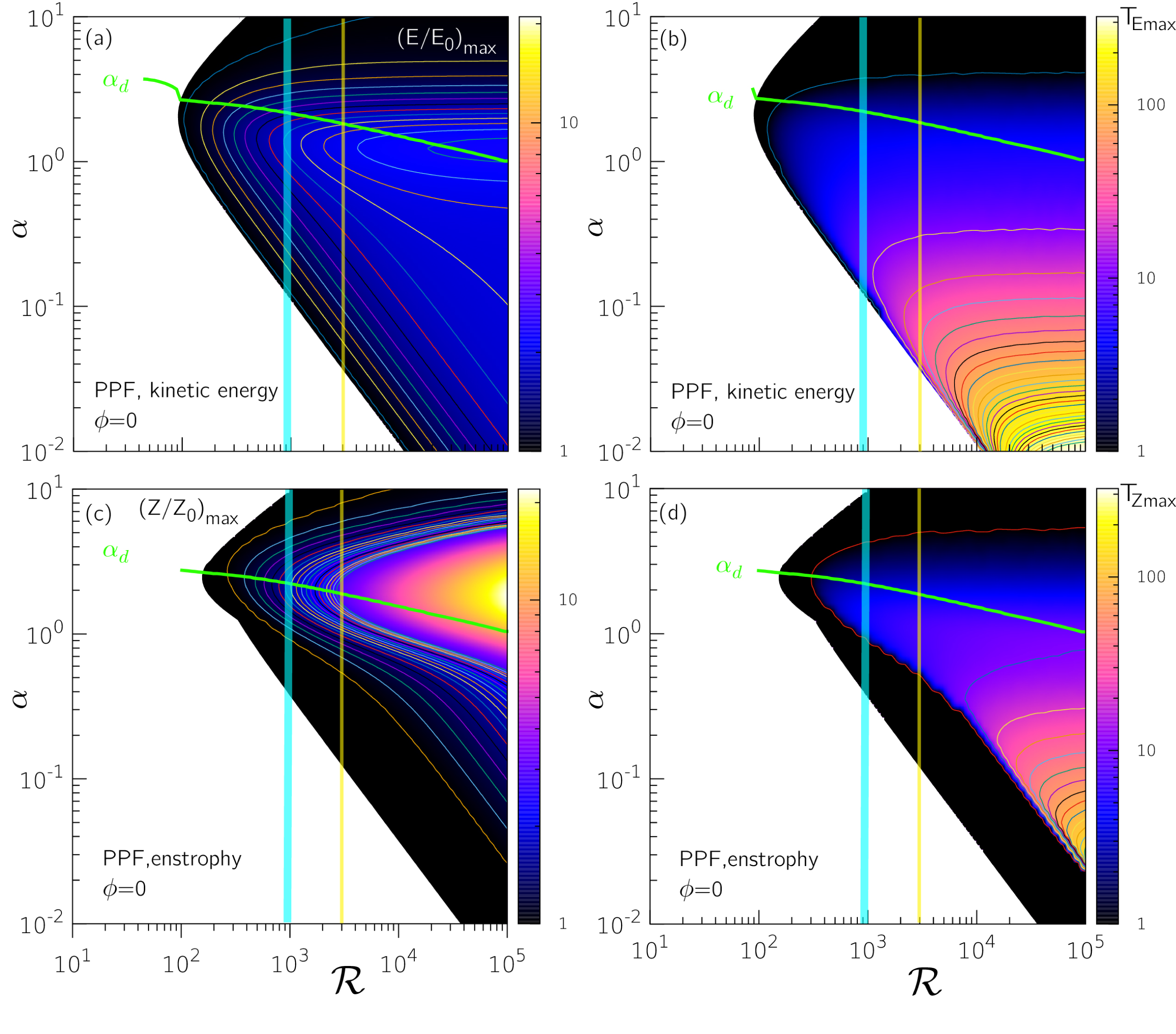}
    \vskip-8pt
    \caption{Maximal kinetic energy and enstrophy achieved during the transient and related time scales, for longitudinal perturbation waves ($\phi=0$) in PPF given by the initial condition of Fig. \ref{fig:initial_v}. Panels (a) and (c)  show the wavenumber-Reynolds number map of maximal transient growth reached during the transient in terms of normalized kinetic energy ($E/E_0$) and enstrophy ($Z/Z_0$), respectively. The time instant corresponding to the maximum energy/enstrophy is shown in the right panels. Each map is built from $3600$ numerical simulations of the Orr-Sommerfeld\textbackslash Squire initial-value problem. The light blue vertical bands represent the global stability threshold, $\mathcal{R}_g$: values collected from experiments in the literature are around $840$ \cite{manneville2016}. In two dimensions, nonlinear analysis of PPF leads to a transitional value of about $2900$ (vertical yellow line) \cite{bayly1988}. The green curve represents the dispersive-to-nondispersive transition between sub-regions \textbf{A} and \textbf{B} in this study. Reproduced from Fig. 5 (b,d) and Fig. SM 1 in Phys. Ref. E. 97, 063102 (2018) [\onlinecite{fraternale2018}]. Copyright American Physical Society.
    } 
\label{fig:maxGZT}
\end{figure*}

According to the representation (\ref{eq:coefficients}), the coefficients of the initial condition are
\begin{equation}
a^0_{q}(y;\alpha,\beta)=\hat q^0 (y;\alpha,\beta)\hat f(\alpha,\beta)\,.
\end{equation}
The velocity-vorticity formulation requires the initial wall-normal velocity $\tilde v^0(y)$ and the initial wall-normal vorticity $\tilde \omega_y^0(y)$. Thus, we specify
\begin{gather}
a^0_{v}(y;\alpha,\beta)=\hat v^0 (y;\alpha,\beta)\hat f(\alpha,\beta)\\
a^0_{\omega_y}(y;\alpha,\beta)=\hat \omega_y^0 (y;\alpha,\beta)\hat f(\alpha,\beta)\,.
\end{gather}
The coefficients of the initial streamwise and spanwise initial components of velocity read
\begin{gather}
a^0_{u}(y;\alpha,\beta)=\hat u^0 (y;\alpha,\beta)\hat f(\alpha,\beta)\\
a^0_{w}(y;\alpha,\beta)=\hat w^0 (y;\alpha,\beta)\hat f(\alpha,\beta)\,,
\end{gather}
and, due to the incompressibility condition and the definition of the velocity curl, they can be derived a posteriori as follows
\begin{gather}
\hat u^0(y;\alpha,\beta)=\frac{i}{k^2}\left[\alpha\dey\hat v^0(y) -\beta\hat\omega_y^0(y)\right]\\
a^0_{w}(y)=\frac{i}{k^2}\left[\beta\dey\hat v^0(y) +\alpha\hat\omega_y^0(y)\right]\,.
\end{gather}
The cross-shear velocity $\hat v^0(y)$ was chosen in order to guarantee a transient growth of the perturbation kinetic energy for any wavenumber included in the packet ($\widehat v^0(y)$ is shown in Fig. \ref{fig:initial_v}). The corresponding maximal growth for longitudinal waves is shown in Fig \ref{fig:maxGZT}. The evolution of this initial condition for individual waves and 2-D wave packets was previously investigated in Ref. \onlinecite{fraternale2018}. This initial velocity profile was obtained from an optimization procedure that seeks the perturbation that maximizes the kinetic energy growth rate  $E^{-1} (\ud E/\ud t)$, at fixed values of $k$ and $\mathcal{R}$. A similar optimization procedure has recently been adopted in the non-modal stability analysis of time-dependent jet flows by \citet{nastro2020}. Such an initial condition excites both symmetric and anti-symmetric Orr-Sommerfeld modes, it is smooth and has a relatively simple shape. Zero initial wall-normal vorticity is set. The resulting initial velocity field is a quadrupole of counter-rotating vortices.
As discussed before, a Gaussian distribution was chosen for $f(x,z)$. This was done in order to represent a smooth, spatially-localized perturbation ($\gamma=0.1$, that is one tenth of the half thickness of the channel) having a similar distribution to disturbances usually introduced in laboratory experiments (where dye is typically injected from a small hole in the wall \cite{carlson1982,klingmann1990,klingmann1992}) or numerical experiments \cite{henningson1993}:
\begin{gather}\label{eq:Gaussian_function}
f(x,z)=\frac{1}{\gamma\sqrt{2\pi}}e^{-[(x-x_0)^2+(z-z_0)^2]/2\gamma^2}\\  \hat{f}(\alpha,\beta)=\mathcal{F}\{f(x,z)\}\,,
\end{gather}
where $\mathcal{F}{}$ denotes the Fourier transform (FT). 
The online movie \texttt{movie\_PPF\_R\_1000\_V.avi} (Multimedia view)  shows the evolution of the wall-normal perturbation velocity $v(x,y=0.6,z)$ for the linear spot shown in Fig. \ref{fig:main_figure}(c) ($\mathcal{R}=1000$, with simulation parameters as reported in Tab. \ref{tab:WPparameters}). The movie \texttt{movie\_waterfall\_PPF\_Re1000\_V.avi} (Multimedia view) shows a pseudo-3D visualization of the wall-normal perturbation velocity for the same simulation.

\subsection{Wave packet spatial spreading rates\label{sec:wave-packet_features}}

The wave packet energy centroid is computed for each squared field component (at a fixed distance from the wall $y_0=-0.6$), and for the $y$-averaged kinetic energy and enstrophy. We use the following notations for the averaging operations over $y$, $z$ and $x$ (that is, for $xz$-, $xy$- and $yz$-reduced quantities, respectively):
\begin{gather}
	\langle \widetilde q \rangle_{y}(t,x,z)=\frac{1}{L_{y}}\int \widetilde q\ud y\,,\\ \nonumber
	\langle \widetilde q \rangle_{z}(t,x,y)=\frac{1}{L_{z}}\int \widetilde q\ud z\,,\\ \nonumber
	\langle \widetilde q \rangle_{x}(t,y,z)=\frac{1}{L_{x}}\int \widetilde q\ud x\,. \label{eq:reduced_quantity}
\end{gather}
The wave packet center of a generic field component $\widetilde q$ is then defined as
\begin{equation}
x_{\mathrm{G}}(t,y_0)=\frac{\int x\langle \widetilde q^2\rangle_{z} \ud x}{\int \langle \widetilde q^2\rangle_{z} \ud x}\,.
\end{equation}
In particular, it can be defined with respect to the average kinetic energy $\langle  E \rangle_y=\langle\widetilde u^2+\widetilde v^2+\widetilde w^2\rangle_{y}$ and enstrophy $\langle  \Omega \rangle_y=\langle \widetilde \omega_{x}^2+\widetilde \omega_{y}^2+\widetilde \omega_{z}^2\rangle_{y}$:
\begin{gather}\label{eq:baricentro}
x_{\mathrm{G}}(t)=\frac{\int \langle \langle  E\rangle_{y}\rangle_{z} x\ud x}{\int \langle\langle   E\rangle_{y}\rangle_{z} \ud x}\,,\hskip20pt
x_{\mathrm{G}}(t)=\frac{\int \langle \langle  \Omega \rangle_{y}\rangle_{z}x \ud x}{\int \langle\langle    \Omega  \rangle_{y}\rangle_{z}\ud x}\,.
\end{gather}

\begin{figure*}
	\includegraphics[width=0.85\textwidth]{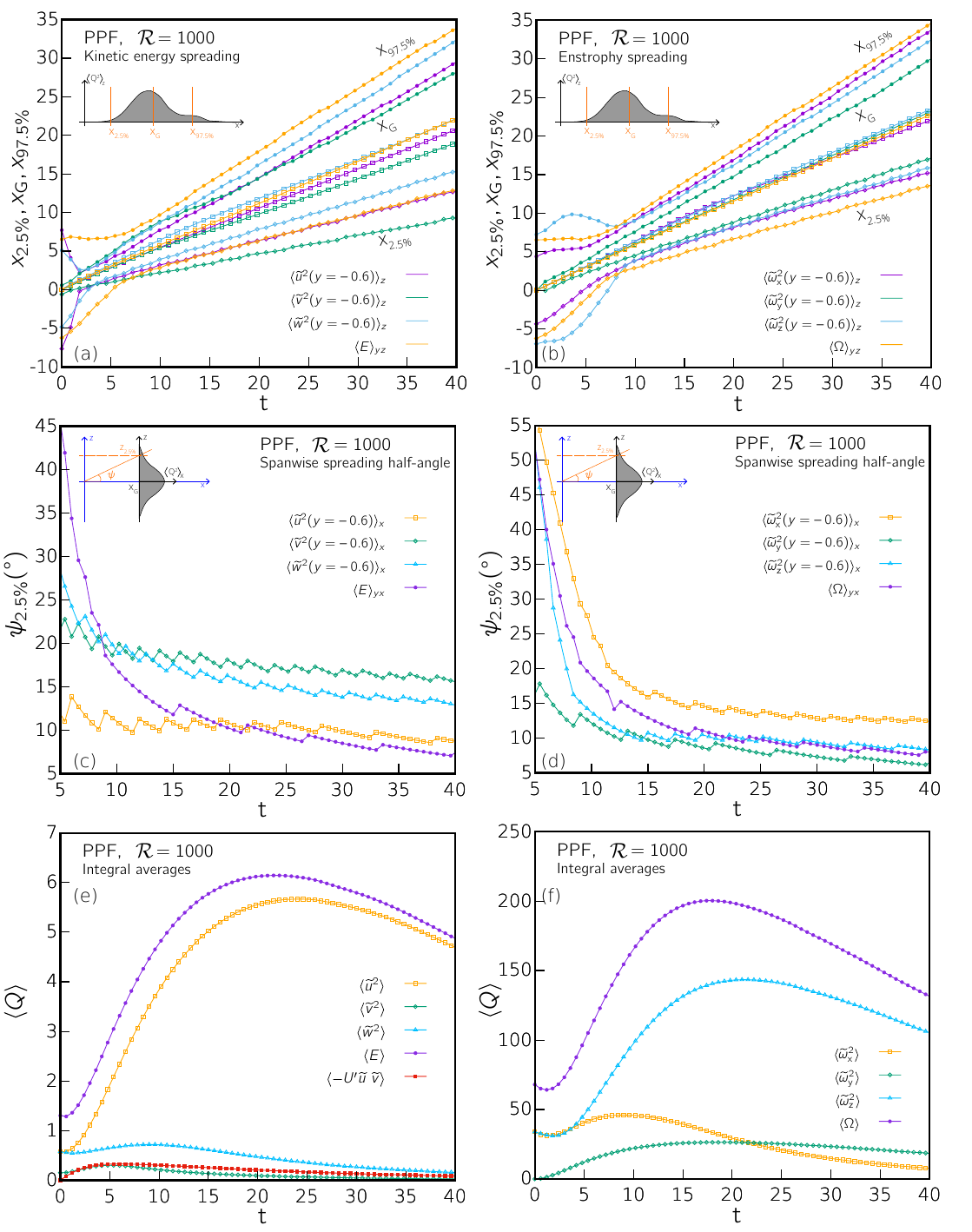}
	\caption{Spreading rates (panels a,b), spreading half-angles (panels c,d), and volume-averaged energy and enstrophy (panels e,f) for the numerical simulation of Fig. \ref{fig:main_figure}(c) [see definitions (\ref{eq:baricentro}), (\ref{eq:x_rear}), (\ref{eq:x_front}), (\ref{eq:lateral_spread}) and (\ref{eq:half_angle}) in App. \ref{sec:wave-packet_features}]. Reproduced from Fig. 28 in F. Fraternale, PhD thesis (2017) [\onlinecite{fraternale_phdthesis}]; licensed under a Creative Commons Attribution (CC BY) license.}
	\label{fig:spreading_rates}
\end{figure*}

The streamwise spreading is computed by seeking the portion of the domain which contains 95\% of the energy. The boundaries of this region are named $x_{\mathrm{rear}}$ and $x_{\mathrm{front}}$, respectively. They are defined as the locations of the 2.5\textsuperscript{th} and 97.5\textsuperscript{th} percentile of the $xy$-reduced, squared variable being considered, respectively: 
\begin{gather}\label{eq:x_rear}
x_{\mathrm{rear}}(t):\hskip10pt \int_{-\infty}^{x_{2.5\%}}\langle \widetilde{q}^2\rangle_{z} \ud x=0.025 \int_{-\infty}^{\infty} \langle\widetilde q^2\rangle_{z} \ud x\,,\\
x_{\mathrm{front}}(t):\hskip10pt\int_{-\infty}^{x_{97.5\%}}\langle \widetilde{q}^2\rangle_{z} \ud x=0.975 \int_{-\infty}^{\infty} \langle\widetilde q^2\rangle_{z} \ud x\,.\label{eq:x_front}
\end{gather}
Analogously,  the spanwise spreading and the spreading half-angle are computed from the $xz$-reduced squared quantities:
\begin{gather}
z_{\mathrm{spread}}(t):\hskip10pt \int_{-\infty}^{z_{2.5\%}}\langle \widetilde{q}^2\rangle_{x} \ud z=0.025 \int_{-\infty}^{\infty} \langle\widetilde q^2\rangle_{x} \ud z\,,\label{eq:lateral_spread}\\
\psi_{\mathrm{spread}}=\tan^{-1}\left(\frac{z_{\mathrm{spread}}}{x_{\mathrm{G}}}\right)\,. \label{eq:half_angle}
\end{gather}
Figure \ref{fig:spreading_rates} presents the spreading rates of a wave packet at $\mathcal{R}=1000$.
Furthermore, Table \ref{tab:WPspreading500} displays the spreading rate data for a $\mathcal{R}=500$ case.
\begin{table}
		\begin{tabular}{l|cccc|cccc} 
		\multicolumn{9}{c}{\hskip0pt {\bf $\mathcal{R}=500$}} \\ 
		&  ${\widetilde u}$  &${\widetilde v}$ &${ \widetilde w}$&${ \langle E\rangle_y}$&$ {\widetilde \omega}_{{x}}$ &$ {\widetilde \omega}_{{y}}$&$ {\widetilde \omega}_{{z}}$&$ {\langle\Omega \rangle}_{{y}}$\\ 	\hline
		${{x}_{\mathrm{G}}}$                   &     0.54       &  0.50     &  0.56 & 0.56&   0.54&0.59&0.59& 0.58\\ 
		${{x}_\mathrm{rear}}$             &    0.33       &   0.27      &  0.41 & 0.36& 0.42& 0.43& 0.43&0.38\\ 
		${{x}_\mathrm{front}}$           &  0.75      &    0.71     &  0.84 & 0.80& 0.84 & 0.78& 0.81 &0.80\\ 
		${{z}_\mathrm{spread}}$             &    0.08       &  0.12       & 0.10  & 0.06 &  0.08&0.06& 0.08&0.05\\ 
		${\psi}{_\mathrm{spread}}$    &   10.5$^\circ$      &     16.1$^\circ$       &    15.1$^\circ$  & 10.2$^\circ$&13.8$^\circ$ & 9.43$^\circ$&9.98$^\circ$ &10.2$^\circ$\\  
	\hline
	\end{tabular}
		\caption[Spreading rates of 3-D wave packets.]{\label{tab:WPspreading500}
		Spreading rates at $\mathcal{R}=500$, time averages computed for $t\in[15,35]$. The spreading rates of the field components are computed at a fixed distance from the wall, $y=-0.6$, while the kinetic energy and the enstrophy are $y$-averaged.}
\end{table}


\providecommand{\noopsort}[1]{}\providecommand{\singleletter}[1]{#1}%
%

\end{document}